\documentclass{article}
\usepackage{dcase2019,amsmath,graphicx,url,times,booktabs, tabularx}
\usepackage{algorithm}
\usepackage{algorithmic}
\usepackage{rotating}
\usepackage{hyperref}
\usepackage{longtable}
\usepackage{caption}
\usepackage{subcaption}
\usepackage{float}
\usepackage{ifpdf}
\usepackage{multirow}
\usepackage{multicol}
\usepackage{amsmath}
\usepackage{amssymb}
\usepackage{microtype}
\title{Guided Learning Convolution System for DCASE 2019 Task 4}

\name{Liwei Lin$^{1,2}$,
       Xiangdong Wang$^{1}$,
       Hong Liu$^{1}$, 
       YueLiang Qian$^{1}$,
       }

 \address{$^1$Bejing Key Laboratory of Mobile Computing and Pervasive Device, \\
 Institute of Computing Technology, Chinese Academy of Sciences, Beijing, China\\
  $^2$University of Chinese Academy of Sciences, Beijing, China\\ 
         \{linliwei17g, xdwang, hliu, ylqian\}@ict.ac.cn\\
  }

\begin{document}

\ninept
\maketitle

\begin{sloppy}

\begin{abstract}
In this paper, we describe in detail the system we submitted to DCASE2019 task 4: sound event detection (SED) in domestic environments. We employ a convolutional neural network (CNN) with an embedding-level attention pooling module to solve it. By considering the interference caused by the co-occurrence of multiple events in the unbalanced dataset, we utilize the disentangled feature to raise the performance of the model. To take advantage of the unlabeled data, we adopt Guided Learning for semi-supervised learning. A group of median filters with adaptive window sizes is utilized in the post-processing of output probabilities of the model. We also analyze the effect of the synthetic data on the performance of the model and finally achieve an event-based F-measure of $45.43\%$ on the validation set and an event-based F-measure of $42.7\%$ on the test set. The system we submitted to the challenge achieves the best performance compared to those of other participates.
\end{abstract}


\section{Introduction}
\label{sec:intro}

DCASE2019 task 4 is the follow-up to DCASE2018 task 4 \cite{Serizel2018}, which aims at exploring the possibility of the large-scale sound event detection using weakly labeled data (without timestamps) and unlabeled data. Different from DCASE2018 task 4, DCASE2019 task 4 introduces an additional strongly annotated synthetic training set.

Sound event detection (SED) consists in recognizing the presence of sound events in the segment of audio and detecting their onset as well as offset. Due to the high cost of manually labeling data, it is essential to efficiently utilize weakly-labeled data and unlabeled data. Simultaneously, the different physical characteristics of events (such as different duration) and the unbalance of the available training set also increase the difficulty of the multi-class SED in domestic environments. For DCASE2019 task4, there are $5$ issues to be resolved:
\begin{itemize}
\item[1)]
How to learn efficiently with weakly-labeled data?
\item[2)]
How to learn efficiently with unbalanced training set?
\item[3)] How to combine weakly-supervised learning with semi-supervised learning efficiently using weakly-labeled data and unlabeled data?
\item[4)]
How to design a better post-processing method on the output probabilities of the model to detect more accurate boundaries according to the characteristics of each event category?
\item[5)]
Does the strongly annotated synthetic training set help?
\end{itemize}

In this paper, we present a system to solve all these five issues. For issue 1 and 2, we utilize convolutional neural network (CNN) with the embedding-level attention pooling module and disentangled feature \cite{lin2019disentangled} to solve them. 
%For issue 2, we focus on the interference caused by the co-occurrence of multiple events in the unbalanced dataset and combine Disentangled Feature with the CNN-MIL framework mentioned above to reduce the impact of it. 
For issue 3, we adopt a semi-supervised learning method named Guided Learning \cite{lin2019you}. For issue 4, according to varied duration of different event categories, we employ a group of median filters with adaptive window sizes in the post-processing of output probabilities of the model.
% we find that the duration of different event categories varies greatly and the duration of the same event categories is similar to each other in real life. Therefore,
% is designed to obtain more accurate event detection boundaries
For issue 5, we simply regard the strongly annotated synthetic training set as a weakly annotated training set and conduct a series of ablation experiments to explore its effects on weakly-supervised learning and unsupervised learning separately.

In the rest of this paper, we introduce our methods in Section~\ref{Methods}, describe in detail our experiments in Section~\ref{Experiments} and draw conclusions in Section~\ref{Conclusion}.

\section{Methods}
\label{Methods}
In this section, we discuss the solution for issue 1 in Section~\ref{Class-wise attention pooling with specialized decision surface}, the solution for issue 2 in Section~\ref{Disentagled feature}, the solution for issue 3 in Section~\ref{Guided learning with a more professional teacher} and the solution for issue 4 in Section~\ref{Adaptive post-processing}.

\subsection{A CNN model with the embedding-level attention pooling module}
\label{Class-wise attention pooling with specialized decision surface}
In this section, we describe in detail the model we employ. As shown in Figure~\ref{fig00}, the model comprises $3$ parts: a feature encoder, an embedding-level attention pooling module and a classifier. The feature encoder encodes the input feature of an audio clip into high-level feature representations. Assuming that there are $C$ event categories to detect, then the embedding-level attention pooling module integrates these high-level feature representations into $C$ contextual representations. Eventually, the clip-level probabilities can be obtained by passing this $C$ contextual representations through the classifier. 

As shown in Figure~\ref{fig01}, the feature encoder we employs is composed of a Batch normalization layer \cite{ioffe2015batch}, 3 Max pooling layers and 3 CNN blocks, each of which consists of a CNN layer, a Batch normalization layer and a ReLU activation layer as shown in Figure~\ref{fig02}. And the classifier for each contextual representation is a fully-connected layer with a Sigmoid activation layer.

The ability of this model to carry out weakly-supervised learning attributes to its embedding-level attention pooling module. Let $\mathbf{x}=\left\{ x_1,...,x_T\right\}$ be the high-level feature representations generated by the feature encoder and $\mathbf{y}=\left\{ y_1,...,y_C\right\}$ ($y_c\in\left\{0,1\right\}$) be the groundtruth, where $T$ denotes the number of total frames of the high-level feature representations. 

Then for each category $c$, the embedding-level attention pooling gives different weights $\mathbf{a_c} = \{a_{c1}, ...,  a_{cT}\}$ to the corresponding $x_t$ in $\mathbf{x}$. Then the contextual representation $\mathbf{h}=\left\{h_1,h_2,...,h_C\right\}$ can be obtained by the following way:
\begin{equation}
  h_c=\sum_t a_{ct}\cdot x_{t}
  \label{eq2}
\end{equation}

\begin{figure}[t]
\begin{center}

\begin{minipage}[minipage1]{0.45\columnwidth}
\begin{subfigure}{\textwidth}
\includegraphics[width=\textwidth]{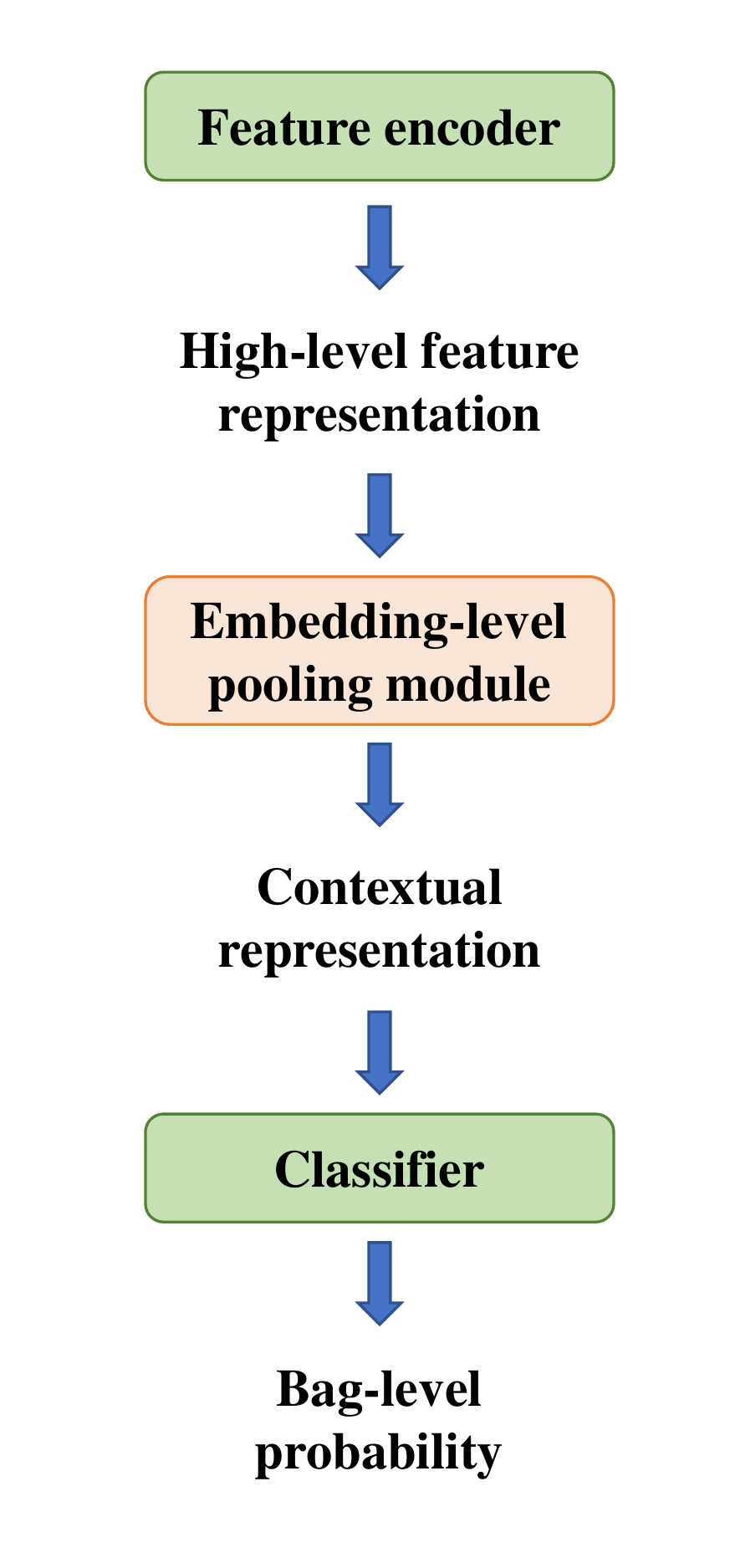}
\caption{Model framework}
\label{fig00}
\end{subfigure}
\end{minipage}
\begin{minipage}[minipage1]{0.45\columnwidth}
\begin{subfigure}{\textwidth}
\includegraphics[width=\textwidth]{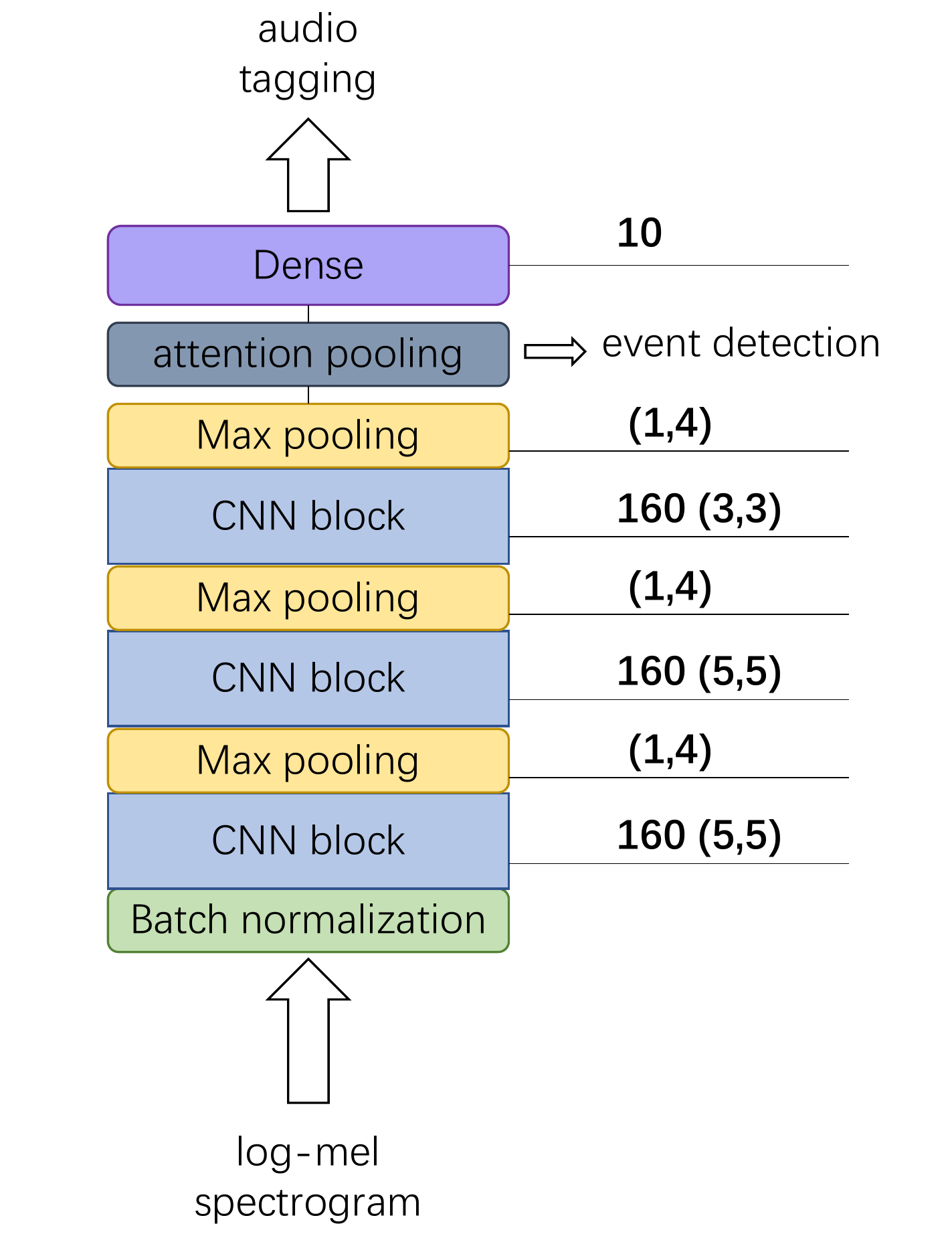}
\caption{Detailed model architecture}
\label{fig01}
\end{subfigure}
\begin{subfigure}{\textwidth}
\includegraphics[width=\textwidth]{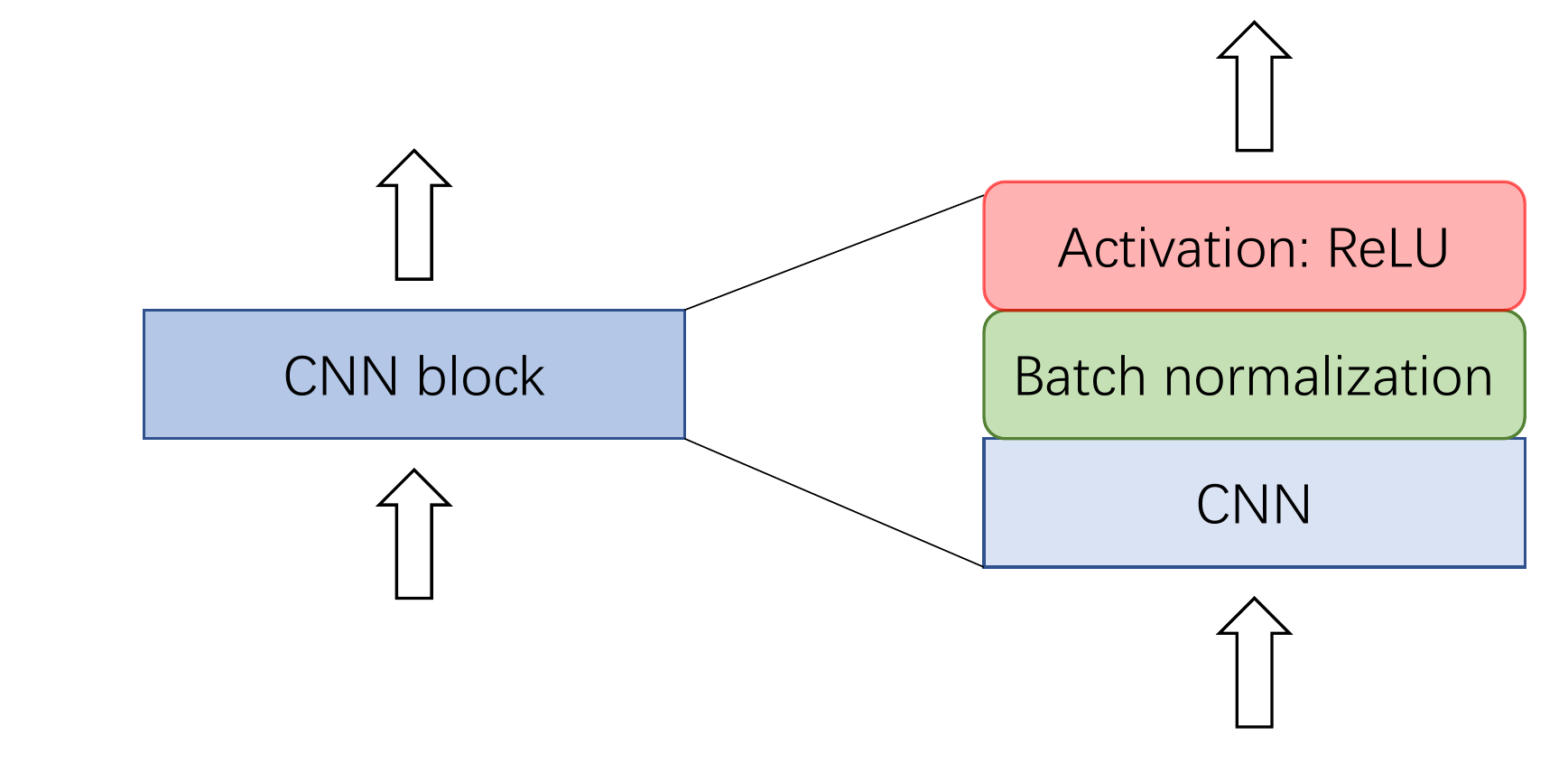}
\caption{CNN block}
\label{fig02}
\end{subfigure}
\end{minipage}
\caption{CNN model with the embedding-level attention pooling module.}\label{fig0}
\end{center}
\vskip -0.2in
\end{figure}
Such an $\mathbf{a_c}$ enables the model to treat each frame differently. Important frame $x_t$ in $\mathbf{x}$ with larger $a_{ct}$ contributes more to $h_c$. The embedding-level attention pooling module generates $\mathbf{a_c}$ by the following way:

\begin{equation}
a_{ct}=\frac{\exp\left( (w_c^{T}x_t+b_c)/d\right)}{\sum_k \exp\left((w_c^{T}x_k+b_c)/d\right)}
\end{equation}
where $d$ is equal with the dimension of $\mathbf{x}$, $w_c^{T}$ is a trainable vector, and $b_c$ is the trainable bias.

More importantly, $a_{ct}$ possess the ability to indicate key frames of an audio and is able to generate frame-level probabilities as explained in \cite{lin2019disentangled}:

\begin{equation}
\hat{p}\left(y_c\mid x_t\right)=
\mathbf{\sigma}\left( w_c^{T} x_t+b_c\right)
\label{eq4}
\end{equation}
where $\mathbf{\sigma}$ is Sigmoid function. 

Assuming that $\mathbf{\hat{P}}\left(y_c\mid\mathbf{x}\right)$ is the clip-level probabilities for event category $c$, then the clip-level prediction is:
\begin{equation}
\mathbf{\phi_{c}}\left(\mathbf{x}\right)=\left\{\begin{matrix}
1,&\mathbf{\hat{P}}\left(1\mid\mathbf{x}\right)\geq \alpha \\ 
0,&otherwise
\end{matrix}\right.
\end{equation}

The frame-level prediction is:

\begin{equation}
\mathbf{\varphi_{c}}\left(\mathbf{x},t\right)
=\left\{\begin{matrix}
1,&\hat{p}\left(1\mid x_t\right)\cdot\mathbf{\phi_{c}}\left(\mathbf{x}\right)\geq \alpha \\ 
0,&otherwise
\end{matrix}\right.
\label{eq3}
\end{equation}

Without loss of generality, we set $\alpha=0.5$.

\begin{figure}[t]
  \centering
  \centerline{\includegraphics[width=\columnwidth]{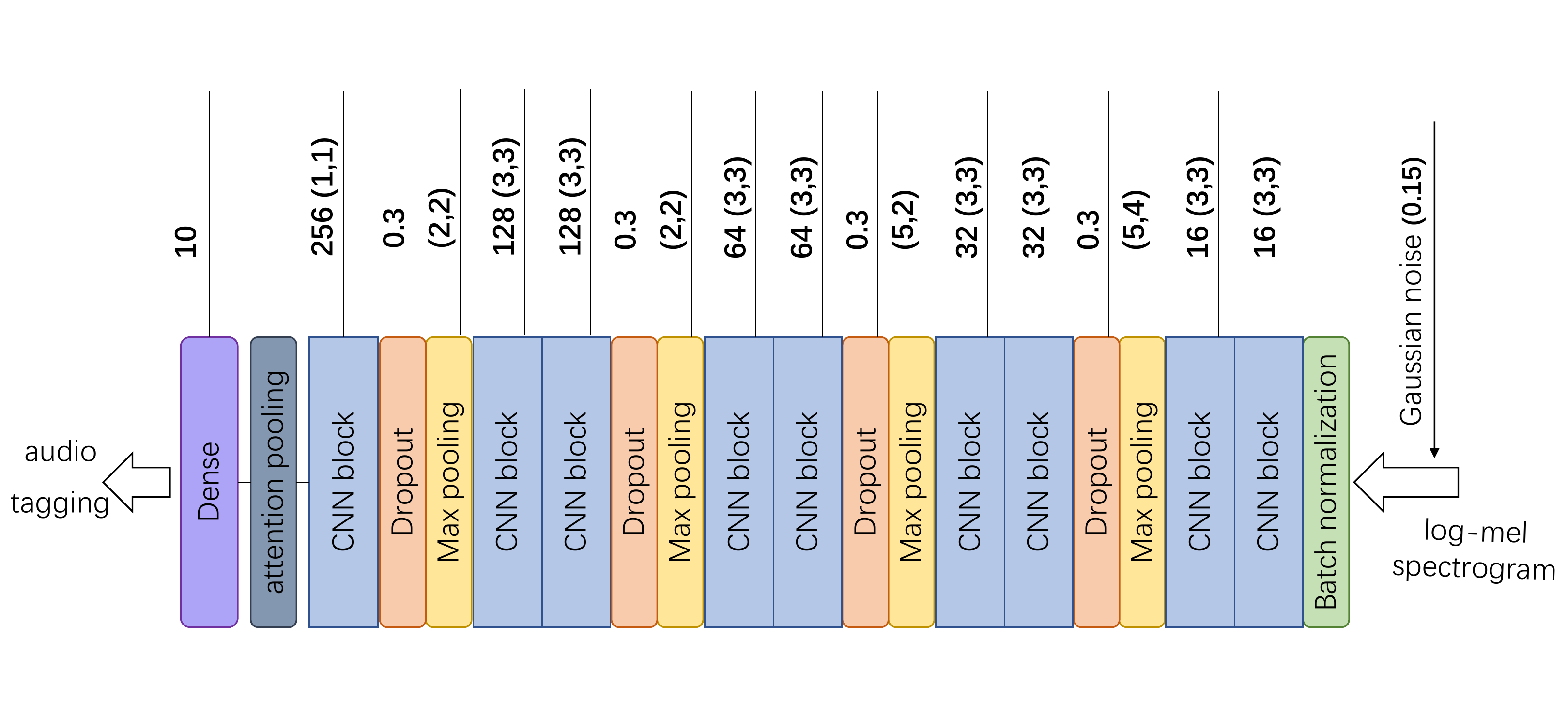}}
  \caption{The model architecture of the PT-model.}
  \label{fig1}
  \vskip -0.1in
\end{figure}

\subsection{Disentangled feature}
\label{Disentagled feature}
We take disentangled feature (DF) \cite{lin2019disentangled}, which re-models the high-level feature subspace of each event category according to the prior information without pre-training, to mitigate the effect of the interference caused by the co-occurrence multiple events.

Assuming that $\mathbf{\chi^d}$ ($\mathbf{x}\subset \mathbf{\chi^d}$) is a d-dimensional space generated by the feature encoder and $\ss=\left\{e_1,e_2,...,e_d\right\}$ is an orthogonal basis of $\mathbf{\chi^d}$ where the element of $e_i$ in $i^{th}$ dimensional is 1. DF selects specific bases of $\mathbf{\chi^d}$ to construct a specific subspace for each category and the basis of the re-modeled feature space $\chi_c^{'}$ of category $c$ is
\begin{equation}
\ss_c^{'} =\left\{e_1,e_2,...,e_{k_c}\right\}
\end{equation}
\begin{equation}
k_c =\lceil\left(\left(1-m\right)\cdot f_c+m\right)\cdot d\rceil
\end{equation}
\begin{equation}
f_c = \sum_{i}^{C}\frac{r_i\cdot {N_c}_i}{R}
\end{equation}
\begin{equation}
R = \underset{c}{\max}\; \sum_{i=1}^{C}r_i\cdot {N_c}_i
\end{equation}
where $m$ is a constant to avoid too-small $k_c$ and ${N_c}_i$ is the number of clips containing $i$ categories including category $c$ in the training set. The constant coefficient $r_i$ denotes the importance these clips: 
\begin{equation}
r_i=\left\{\begin{matrix}
 1,& i=1\\ 
 0,& otherwise
\end{matrix}\right.
\label{eq19}
\end{equation}

\subsection{Guided Learning}
\label{Guided learning with a more professional teacher}

To combine weakly-supervised learning with semi-supervised learning, we utilize Guide Learning (GL) proposed in \cite{lin2019you} with a more professional teacher model (PT-model) to guide a more promising student model (PS-model).

\begin{algorithm}[t]
   \caption{Guided learning pseudocode.}
   \label{algorithm1}
\begin{algorithmic}
\REQUIRE $x_k = $ training input with index $k$ 
\REQUIRE $L = $ set of weakly-labeled training input 
\REQUIRE $U = $ set of unlabeled training input 
\REQUIRE $y_k = $ label of weakly-labeled input $x_k\in L$
\REQUIRE $S_\theta\left(x\right)=$ neural network of the PS-model with trainable parameters $\theta$ 
\REQUIRE $T_{{\theta}^{'}}\left(x\right)=$ neural network of the PT-model model with trainable parameters ${\theta}^{'}$ 
\REQUIRE $g\left(x\right)=$ stochastic input augmentation function 
\REQUIRE $J\left(t,z\right)=$ loss function 
\REQUIRE $\mathbf{\phi}(z)=$ prediction generation function
\ENSURE {$\theta, {\theta}^{'}$}
\FOR{$i=1 \rightarrow num\_epoches$}
\IF{$i>start\_epoch$}
\STATE $a\leftarrow 1-{\gamma}^{i-start\_epoch}$ \hfill$\triangleright$ calculate the weight of unsupervised loss of the PT-model
\ELSE
\STATE $a\leftarrow 0$
\ENDIF
\FOR{each minibatch $\mathbf{\ss}$}
\STATE $s_k\leftarrow S_\theta\left(x_k\in \mathbf{\ss}\right)$\hfill$\triangleright$ the coarse-level predicted probability of the PS-model
\STATE $t_k\leftarrow T_{{\theta}^{'}}\left(g(x_k)\in \mathbf{\ss}\right)$\hfill$\triangleright$ the coarse-level predicted probability of the PT-model
\STATE $\tilde{s_k}\leftarrow \mathbf{\phi}\left(s_k\right)$
\hfill$\triangleright$ convert the predicted probability into 0-1 prediction
\STATE $\tilde{t_k}\leftarrow \mathbf{\phi}\left(t_k\right)$
\IF{$x_k \in L$}
\STATE $loss\leftarrow\frac{1}{\left |\mathbf{\ss}\right |} \left\{
\sum_{x_k \in {J \cap \mathbf{\ss}}} \left[J\left(y_k,s_k\right)+
J\left(y_k,t_k\right)\right]\right\}$
\ENDIF
\IF{$x_k \in U$}
\STATE $loss\leftarrow\frac{1}{\left |\mathbf{\ss}\right |} \left\{
\sum_{x_k \in {U \cap \mathbf{\ss}}} \left[J\left(\tilde{t_k},s_k\right)+
a\cdot J\left(\tilde{s_k},t_k\right)\right]\right\}$
\ENDIF
\STATE update $\theta, {\theta}^{'}$\hfill$\triangleright$ update network parameters
\ENDFOR
\ENDFOR
\vskip -0.2in
\end{algorithmic}
\end{algorithm}

The architecture of the PS-model is consistent with the model described in Section~\ref{Class-wise attention pooling with specialized decision surface} and we show that of the PT-model in Figure~\ref{fig1}. The CNN feature encoder of the PT-model is considered to be better designed than the PS-model on the audio tagging performance with larger sequential sampling size and less trainable parameters. This is because that the larger sequential sampling size allows the CNN feature encoder of the PT-model to have a larger receptive field followed by better exploitation of contextual information.

However, the larger sequential sampling size also disables the PT-model to see finer information due to the compress of sequential information. Therefore, the PS-model is designed with smaller sequential sampling size to see finer information and then achieves better frame-level prediction.

This gap between their ability makes it possible to optimize the PS-model with the guide of the PT-model using unlabeled data. As shown in Algorithm~\ref{algorithm1}, an end-to-end process is employed to train these two models.

\subsection{Adaptive post-processing}
\label{Adaptive post-processing}

The median filter is utilized for post-processing of the frame-level probabilities output by the model. Instead of determining the window size of the median filter empirically, we adopt a group of median filters with adaptive window sizes for different event categories by the following formulation based on the varying length of different event categories in real life:
\begin{equation}
  \label{eq1}
    S_{win}=duration_{ave}\cdot\beta
\end{equation}

All the frame-level probabilities output by the network are smoothed by a group of median filters with these adaptive window sizes. After smoothed, the probabilities are converted into the 0-1 prediction with a threshold of $0.5$ as described in Section~\ref{Class-wise attention pooling with specialized decision surface}. Then the operation of smoothing is repeated again on the final frame-level prediction.

%\begin{figure}[t]
%\begin{center}

%\begin{minipage}[minipage1]{0.45\columnwidth}
%\begin{subfigure}{\textwidth}
%\includegraphics[width=\textwidth]{fi2.pdf}
%\caption{PT-model}
%\end{subfigure}
%\end{minipage}
%\begin{minipage}[minipage1]{0.45\columnwidth}
%\begin{subfigure}{\textwidth}
%\includegraphics[width=\textwidth]{fi3.pdf}
%\caption{PS-model}
%\end{subfigure}
%\begin{subfigure}{\textwidth}
%\includegraphics[width=\textwidth]{fi4.pdf}
%\caption{CNN block}
%\end{subfigure}
%\end{minipage}
%\caption{The model architectures of PT-model and PS-model.}\label{fig1}
%\end{center}
%\vskip -0.2in
%\end{figure}

\section{Experiments}
\label{Experiments}
\subsection{DCASE 2019 Task 4 Dataset}
\label{Dataset}
The dataset \cite{Turpault2019,Gemmeke2017audioset,Fonseca2017freesound,font2013freesound,Dekkers2017} of DCASE2019  task 4 is divided into 4 subsets: the weakly annotated training set (1578 clips), the unlabeled training set (14412 clips), the strongly annotated validation set (1168 clips) and the strongly annotated synthetic training set (2045 clips) \cite{salamon2017scaper}.
%The striking features of this dataset is unbalance. Each 10-second audio clip in the dataset contains one or more (or None) 10 events.
%of the following : Alarm ( bell or ringing), Blender, Cat, Dishes, Dog, Electric shaver (or toothbrush), Frying, Running water, Speech, Vacuum cleaner.
We integrate the weakly annotated training set, the unlabeled training set and the strongly annotated synthetic training set (actually we only use weakly labels during training) into a training set and take the validation set as our validation set. The average duration of each event category in the synthetic set is shown in Figure~\ref{fig2}.

\subsection{Feature exaction and post-processing}
\label{Feature exaction}
We produce 64 log mel-bank magnitudes which are extracted from 40 ms frames with $50\%$ overlap ($n_{FFT}=2048$) using librosa package \cite{brian_mcfee-proc-scipy-2015}. All the 10-second audio clips are extracted to feature vectors with $500$ frames. In post-processing, we take $\beta=\frac{1}{3}$ (see details in Section~\ref{Adaptive post-processing}) in our experiments and the window sizes for different events are shown in Table~\ref{table2}.

\begin{figure}[t]
  \centering
  \centerline{\includegraphics[width=0.96\columnwidth]{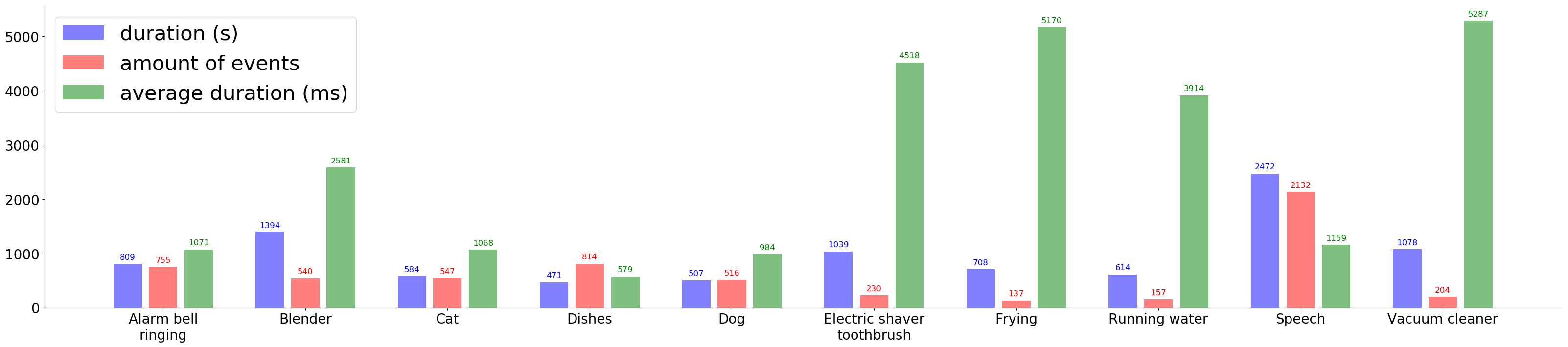}}
  \caption{The total duration, number of events and average duration per event category in the synthetic training set.}
  \label{fig2}
\end{figure}
\begin{figure}[t]
  \centering
  \centerline{\includegraphics[width=0.96\columnwidth]{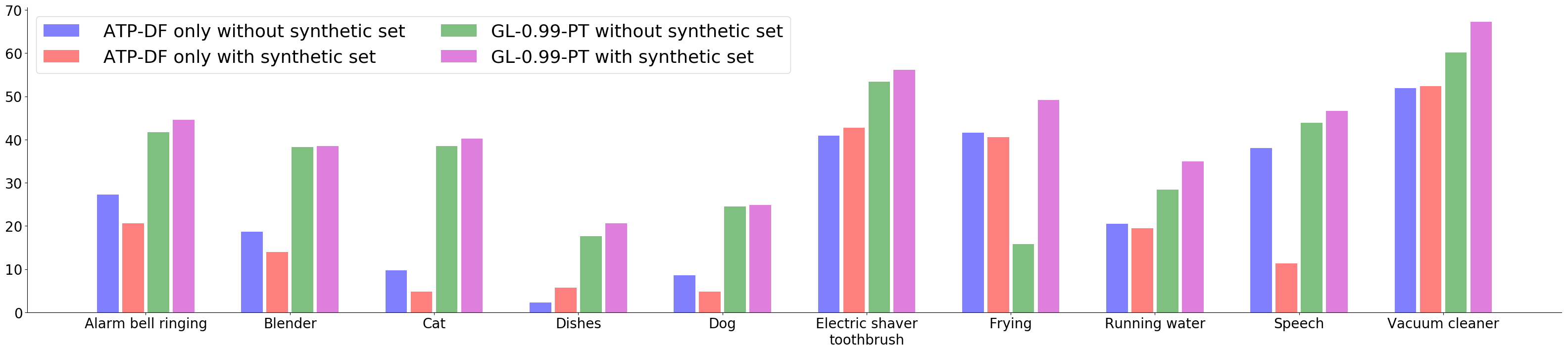}}
  \caption{The class-wise $F_1$ performance}
  \label{fig3}
  \vskip -0.1in
\end{figure}
\subsection{Model architecture}
\label{Model architecture}

The model architectures of the PS-model and PT-model are described in detail in Section~\ref{Methods}. We take $m=0.04$ for DF. The dimension of DF per category is shown in Table~\ref{table2}. The PS-model has about 2.6 times the number of trainable parameters as the PT-model (877380 / 332364). The start epoch for GL is set to $5$. The PS-model with only weakly-supervised learning is named ATP-DF and the co-teaching of the PS-model and the PT-model is named GL-$\alpha$-PT in the performance report, where $\alpha$ is a hyper-parameter for GL discussed in Algorithm~\ref{algorithm1}.

\subsection{Training}
\label{Training}

\begin{table}[t]
  \caption{The dimension of the disentangled feature when $m=0.04$ and the window sizes of the median filters when $\beta=\frac{1}{3}$.}
  \vskip 0.01in
  \label{table2}
  \centering
\begin{tabular}{c|cc}
\hline
\textbf{Event}&\textbf{DF}&\textbf{Window size}\\
&\textbf{dimension}&\textbf{(frame)}\\
\hline
Alarm/bell/ringing&137&17\\
Blender&94&42\\
Cat&134&17\\
Dishes&69&9\\
Dog&132&16\\
Electric shaver/toothbrush&76&74\\
Frying&34&85\\
Running water&160&64\\
Speech&30&18\\
Vacuum cleaner&113&87\\
\hline
\end{tabular}
\vskip -0.1in
\end{table}
\begin{table}[t]
  \caption{The performance of models from top1 and the ensemble of models.}
  \vskip 0.01in
  \label{table6}
  \centering
\begin{tabular}{lcc}
\toprule
\textbf{Model}&\multicolumn{2}{c}{\textbf{Macro $F_1$ (\%)}}\\
&Event-based&Segment-based\\
\midrule
Top1&$44.47$&$66.74$\\
%Top2&$44.02$&$67.07$\\
%Top3&$44.01$&$67.38$\\
%Top4&$43.94$&$67.39$\\
%Top5&$43.80$&$67.99$\\
%Top6&$43.73$&$68.44$\\
Ensemble (Top1-6)&$45.28$&$\mathbf{69.06}$\\
Ensemble (Top2-6)&$\mathbf{45.43}$&$69.02$\\
\bottomrule 
\end{tabular}
\vskip -0.15in
\end{table}
The Adam optimizer \cite{kingma2014adam} with learning rate of 0.0018 and mini-batch of 64 10-second patches is utilized to train models. The learning rate is reduced by $20\%$ per 10 epochs.  Training early stop if there is no more improvement on clip-level macro $F_1$ within 20 epochs. All the experiments are repeated $30$ times and we report the average results. Event-based measures \cite{Mesaros2016_MDPI} with a 200ms collar on onsets and a 200ms / $20\%$ of the events length collar on offsets are calculated over the entire test set.
\subsection{Results}
\label{Results}
As shown in Table~\ref{table3}, GL-0.99-PT (with synthetic set) achieves the best average performance on event-based macro $F_1$. The class-wise $F_1$ performance per event category is shown in Figure~\ref{fig2}. As shown in Table~\ref{table6}, the ensemble of the models (GL-0.99-PT) from top2 to top6 achieves the best performance, improving the performance by 21.73 percentage points from the baseline.
\begin{table}[t]
  \caption{The performance of models}
  \vskip 0.01in
  \label{table3}
  \centering
\begin{tabular}{lcc}
\toprule
\textbf{Model}&\multicolumn{2}{c}{\textbf{Macro $F_1$ (\%)}}\\
&Event-based&Segment-based\\
\midrule
baseline&23.7&55.2\\
\midrule
\multicolumn{2}{l}{without the synthetic training set}&\\
ATP-DF&$25.95\pm3.22$&$56.82\pm1.34$\\
GL-1-PT&$35.19\pm3.86$&$61.14\pm3.14$\\
GL-0.996-PT&$\mathbf{36.50\pm3.71}$&$\mathbf{62.03\pm3.25}$\\
GL-0.99-PT&$36.21\pm4.63$&$61.25\pm2.77$\\
GL-0.98-PT&$33.78\pm2.95$&$57.54\pm3.42$\\
\midrule
\multicolumn{2}{l}{with the synthetic training set}&\\
ATP-DF&$21.65\pm2.55$&$57.02\pm1.93$\\
GL-1-PT&$41.03\pm2.98$&$65.58\pm2.84$\\
GL-0.996-PT&$42.02\pm3.29$&$\mathbf{66.62\pm1.82}$\\
GL-0.99-PT&$\mathbf{42.32\pm2.21}$&$65.78\pm2.63$\\
GL-0.98-PT&$41.16\pm2.42$&$63.89\pm2.20$\\
\bottomrule 
\end{tabular}
\vskip -0.1in
\end{table}
\begin{figure}[t]

  \centering
  \centerline{\includegraphics[width=0.7\columnwidth]{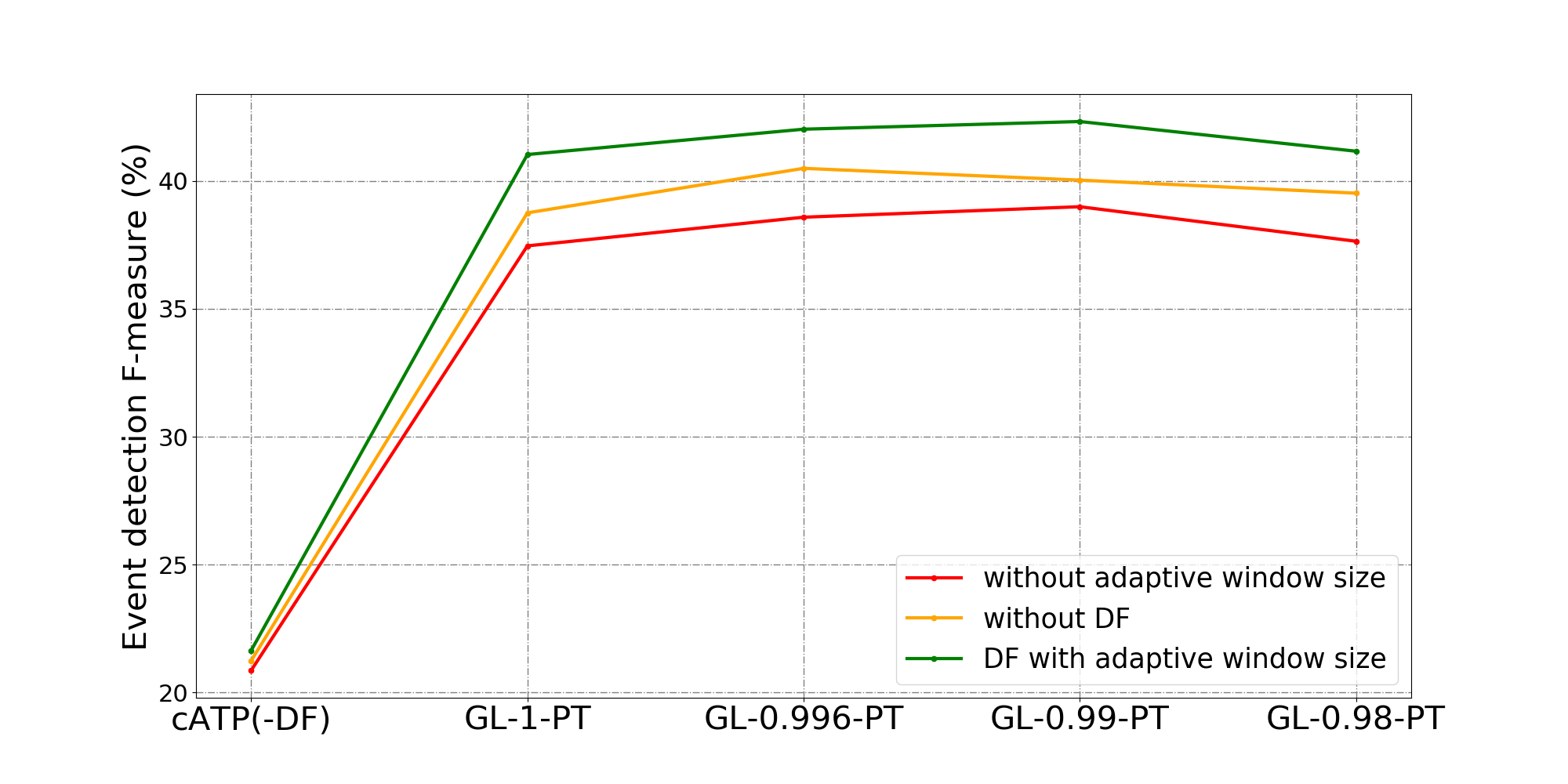}}
  \caption{The performance of models without the disentangled feature (without DF) or smoothed by the median filter with a fixed window size of $27$ (without adaptive window size).}
  \label{fig4}
\vskip -0.2in
\end{figure}
As shown in Table~\ref{table3}, all the models with semi-supervised learning outperform those only with weakly-supervised learning significantly and the model with the best average performance improves the performance by 20.67 percentage points from the weakly-supervised only method. As shown in Figure~\ref{fig4}, the performances of all the models without disentangled feature or adaptive window sizes are poorer than those which has. 

\subsubsection{Does the synthetic training set help?}
\label{Does the synthetic data help?}

As shown in Table~\ref{table3}, when learning only with weakly labeled data, the synthetic training set not only does not help improve the results but also brings negative effects. But when combining weakly-supervised learning with semi-supervised learning, the synthetic training set contributes a lot so that the performance is raised by about 5-8 percentage points. We argue that the model tends to be overfitting in the synthetic training set and have difficulty in recognizing the audio clips from the real-life recording since the number of audio clips in the synthetic training set is almost 1.3 times as much as that in the weakly annotated training set. However, the large scale of unlabeled data complements this weakness and enable the synthetic training set to play a positive role during training.

\subsubsection{Challenge results}
\label{Challenge results}
The model (Ensemble Top1-6) achieves an F-measure of $42.7\%$ on the test set and won the first price in the challenge, which is $0.6$ percentage point ahead of the second place and $0.7$ percentage points ahead of the third place. We note that according to the supplementary metrics released by challenge official, our model achieves the best performance on the Youtube dataset but shows a poorer performance than the second place and the third on the Vimeo dataset. This might be because most of the audios in the dataset are from Youtube. From this point, we guess the data augmentation such as pitch shifting and time stretching might help a lot.

\section{Conclusions}
\label{Conclusion}
In this paper, we present a system for DCASE2019 task 4. Actually, we present a complete system for large-scale weakly labeled semi-supervised sound event detection in domestic environments. We broke the task down into $4$ small sub-problems and came up with solutions for each. We release the implement to reproduce our system at https://github.com/Kikyo-16/Sound\_event\_detection. We employ a CNN model with an embedding-level attention module to carry out weakly-supervised learning and utilize GL to carry out semi-supervised learning. DF is employed to raise the performance of the model by reducing the interference caused by the co-occurrence of multiple event categories. In addition, adaptive post-processing is proposed to get more accurate detection boundaries. We also analyze the effect of the synthetic training set. As a result, we achieve state-of-the-art performance on the dataset of DCASE2019 task4.

\section{ACKNOWLEDGMENT}
This work is partly supported by Beijing Natural Science Foundation (4172058).
% -------------------------------------------------------------------------
% Either list references using the bibliography style file IEEEtran.bst
\bibliographystyle{IEEEtran}
\bibliography{refs}
%
% or list them by yourself
% \begin{thebibliography}{9}
% 
% \bibitem{dcase2016web}
%   \url{http://www.cs.tut.fi/sgn/arg/dcase2016/}.
%
% \bibitem{IEEEPDFSpec}
%   {PDF} specification for {IEEE} {X}plore$^{\textregistered}$,
%   \url{http://www.ieee.org/portal/cms_docs/pubs/confstandards/pdfs/IEEE-PDF-SpecV401.pdf}.
%
% \bibitem{PDFOpenSourceTools}
%   Creating high resolution {PDF} files for book production with 
%   open source tools, 
%   \url{http://www.grassbook.org/neteler/highres_pdf.html}.
%
% \bibitem{eWilliams1999}
% E. Williams, \emph{Fourier Acoustics: Sound Radiation and Nearfield Acoustic
%   Holography}. London, UK: Academic Press, 1999.
% 
% \bibitem{ieeecopyright}
%   \url{http://www.ieee.org/web/publications/rights/copyrightmain.html}.
%
% \bibitem{cJones2003}
% C. Jones, A. Smith, and E. Roberts, ``A sample paper in conference
%   proceedings,'' in \emph{Proc. IEEE ICASSP}, vol. II, 2003, pp. 803--806.
% 
% \bibitem{aSmith2000}
% A. Smith, C. Jones, and E. Roberts, ``A sample paper in journals,'' 
%   \emph{IEEE Trans. Signal Process.}, vol. 62, pp. 291--294, Jan. 2000.
% 
% \end{thebibliography}

\end{sloppy}
\end{document}